\documentclass[a4paper]{article}

\usepackage{cite}
\usepackage[flushleft]{threeparttable}
\usepackage{inputenc}
\usepackage{changepage}
\usepackage{subcaption}
\usepackage{dirtytalk}
\usepackage{caption,tabularx,booktabs}
\usepackage{hyperref}
\usepackage{diagbox}
\usepackage{afterpage}
\usepackage[margin=1in]{geometry}
\usepackage{fancyhdr}
\usepackage{lastpage}
\usepackage[square,sort,comma,numbers]{natbib}
\usepackage{amsmath}
\usepackage{amsfonts}
\usepackage{amssymb}
\usepackage{graphicx}
\usepackage{bm}

\begin{document}

\begin{titlepage}\centering
\vspace*{\fill}
\LARGE A Role for Prior Knowledge in Statistical Classification of the Transition from MCI to Alzheimer's Disease\\
\

\vspace*{\fill}
\begin{center}
\large
Zihuan Liu$^1$, Tapabrate Maiti$^1$ \\
\small  {
$^1$Department of Statistics, Michigan State University, East Lansing, MI, USA}\\
\large
{Andrew R.Bender$^2$}\\
\small {
$^2$Department Epidemiology and Biostatistics and Department of Neurology and Ophthalmology, College of Human Medicine, Michigan State University, East Lansing, MI, USA\\
arbender@msu.edu\\
(517) 432-9277}\\
\vspace*{\fill}
\end{center}
\end{titlepage}

\normalsize
\newpage 
\begin{center}
\section*{Abstract}
\end{center}
The transition from mild cognitive impairment (MCI) to Alzheimer’s disease (AD) is of great interest to clinical researchers. This phenomenon also serves as a valuable data source for quantitative methodological researchers developing new approaches for classification. However, the growth of machine learning (ML) approaches for classification may falsely lead many clinical researchers to underestimate the value of logistic regression (LR), yielding equivalent or superior classification accuracy over other ML methods. Further, in applications with many features that could be used for classifying the transition, clinical researchers are often unaware of the relative value of different selection procedures. In the present study, we sought to investigate the use of automated and theoretically-guided feature selection techniques, and as well as the $L_1$ norm when applying different classification techniques for predicting conversion from MCI to AD in a highly characterized and studied sample from the Alzheimer's Disease Neuroimaging Initiative (ADNI). We propose an alternative pre-selection technique that utilizes an efficient feature selection based on clinical knowledge of brain regions involved in AD. The present findings demonstrate how similar performance can be achieved using user-guided pre-selection versus algorithmic feature selection techniques. Finally, we compare the performance of a support vector machine (SVM) with that of logistic regression on multi-modal data from ADNI. The present findings show that although SVM and other ML techniques are capable of relatively accurate classification, similar or higher accuracy can often be achieved by LR, mitigating SVM's necessity or value for many clinical researchers.

{\it Key Words: Alzheimer’s Disease; logistic regression; mild cognitive impairment; support vector machine. }
\newpage 

\section*{Introduction}
	Alzheimer’s disease (AD) is a progressive, age-related, neurodegenerative disease and the most common cause of dementia ~\cite{bib1,bib2,bib3}. Behaviorally, AD is commonly preceded by mild cognitive impairment (MCI), a syndrome characterized by declines in memory and other cognitive domains that exceed cognitive decrements associated with normal aging ~\cite{bib2,bib23}. However, the prodromal symptoms of MCI are not prognostically deterministic: individuals with MCI tend to progress to probable AD at a rate of $8\%$-$15\%$ per year, and most conversions occur within 3 years of presentation ~\cite{bib4,bib17,bib18}. Research efforts to provide new insights into the incidence of MCI-to-AD conversion have focused largely on clinically or biologically relevant features (i.e., neuroimaging markers, clinical exam data, neuropsychological test scores) and on different methods for statistical classification ~\cite{bib6}. \par
For clinical researchers, however, there may be a tendency to conflate more sophisticated, novel analytic approaches and the value of multimodal information from neuroimaging and clinical assessment. Moreover, whereas statisticians may inherently understand the comparability of different quantitative approaches, the novelty of both big data and data-driven approaches for studying MCI-to-AD conversion may lead clinical researchers to assume that such data-driven methods are inherently superior to more theoretically-grounded approaches. Thus, the value of using extant findings and domain expertise to help guide and constrain the application of newer data-driven approaches capable of capitalizing on emergent big data may be a particularly important consideration for clinical researchers. \par
Statistical classification in clinical research has traditionally utilized binary logistic regression (LR). However, key attributes of modern clinical and neuroimaging data, including high dimensionality and the presence of ground truth estimates of pathology and diagnosis, provide new opportunities for quantitative research. This has led to a substantial expansion in the use of data from the Alzheimer’s Disease Neuroimaging Initiative (ADNI; http://adni.loni.usc.edu) for quantitative research and methodological development, particularly by researchers utilizing and developing prediction and classification methods in machine learning (ML). Besides LR, support vector machine (SVM) has quickly become the most common type of ML classifier for diagnostic prediction and classification with ADNI data. In addition to LR and SVM, deep neural network approaches also offer benefits ~\cite{bib22,bib25}, but have not had the extent of application in ADNI data as SVM and LR. In general, LR works well when the data is linearly separable and the number of data is greater than the number of features, whereas SVM with Gaussian Kernel is mostly used when the data is not linearly separable. Moreover, SVM and LR have similar misclassification rates (MCRs) when used to diagnose malignant tumors from imaging data ~\cite{bib11,bib13}. \par
Indeed, before the rapid expansion of ML research and applied work over the past decade, many clinical researchers and those outside of engineering and mathematically intensive disciplines had little exposure to classification approaches other than LR. Despite its growing popularity, the relative benefits of SVM or other forms of ML~\cite{bib47,bib48} over LR for such classification are not always apparent. Although this may be of little surprise to statisticians and quantitative researchers, such perspectives are often lost on clinical researchers, whose implicit beliefs in the superiority of ML is driven by the volume of publications, rather than through training or empirical demonstration. \par
Most efforts to develop new classification methods for prediction of MCI-to-AD conversion are well suited to integrate measures from multiple sources such as demographics, clinical rating scores, neuropsychological testing, neuroimaging, genetic markers, etc. However, identifying which combination of features most accurately classifies conversion from MCI to AD is a key challenge for ADNI, and may vary by method. The $L_1$ norm regularization method (i.e., $L_1$) is a highly used feature selection technique for LR and SVM. $L_1$ is popular for addressing circumstances in which the number of features is greater than the sample size (i.e., small n, large p) and has been implemented in ADNI data with LR ~\cite{bib8}. Furthermore, $L_1$ and other algorithmic feature selection methods used in ML suffer from one key limitation: they are agnostic to theoretical considerations, and as such, they cannot interpret why selected features are meaningful and important to the model. When sampling from a large pool of features, the algorithmic approaches fail to consider prior knowledge of features and their associations with the relevant systems in variable selection. Therefore, domain expertise and prior knowledge may afford additive or differential value for choosing features and interpreting model results over algorithmic feature selection methods alone.\par
However, most real-world problems occur in the context of additional information about each potential feature and its conceptual relationship with the phenomenon being classified. Other than using $L_1$ feature selection, manually trimming the list of potential predictor variables can also protect against over-fitting, and also offers potential insight into why selected features are important to the model. When guided by prior knowledge, user-guided or ‘manual’ feature selection may be a valuable additional step to help minimize potentially spurious effects. This perspective is frequently lost on applied researchers, as most commonly used variable selection algorithms are context-free – that is, they only look at relationships within the data set, and cannot factor in the wider meanings of variables. Furthermore, this also means that automated algorithms may identify relationships among a large number of predictor variables that are spurious and are unlikely to generalize outside the data set. Although there are a vast number of potential neuroimaging features in ADNI data, the present study focused only on regional brain volumes segmented from structural magnetic resonance imaging (MRI) data, the most commonly used neuroimaging datatype for classifying MCI-to-AD conversion. In contrast to prior studies that used a limited set of volumetric brain features, the present study utilized data generated by modern multi-atlas segmentation methods and analyses included up to 259 features - anatomically specific gray and white matter volumes.\par
The present study addressed two questions regarding commonly used classification approaches for predicting MCI-to-AD conversion in multi-modal data from ADNI. First, we compared performance accuracy of binary LR with SVM in classifying MCI-to-AD conversion. Second, we asked if applying prior knowledge in feature selection outperforms algorithmic variable selection alone. We hypothesized that 1) LR would perform comparably to SVM, and 2) user-guided variable selection would outperform algorithmic variable selection alone. This work is intended to demonstrate to clinical researchers the benefit of using ML in an informed fashion, rather than as a ‘black box’ that obscures clear interpretation. Moreover, we wish to emphasize that this study is not meant to highlight a novel innovation in quantitative methods, but rather to provide an important example to applied researchers regarding the comparable value of ML methods and importance of domain expertise in classification with ADNI data. \par

\section*{Materials and Data}

The data used in the preparation of this study were obtained from the Alzheimer’s Disease Neuroimaging Initiative (ADNI). ADNI is an ongoing joint public-private effort to utilize neuroimaging, other biological markers, and clinical and neuropsychological assessment to measure the incidence and progression of MCI to early AD. Determination of sensitive and specific markers of very early AD progression is intended to aid researchers and clinicians to develop new treatments and monitor their effectiveness, as well as lessen the time and cost of clinical trials. Data in the present study came from all sites across the U.S and Canada. All ADNI subjects were between 55 and 90 years old, spoke English or Spanish as their native language, and had a study partner able to provide an independent assessment of functioning. 
\par

This study used a subset of the MCI subjects from ADNI-1 who had data from demographic, clinical cognitive assessments, APOE4 genotyping, and MRI measurements. In total, there are 819 individuals with a baseline diagnosis of MCI. To evaluate differences in classification due to participant inclusion and drop out, we subdivide the sample into two overlapping groups. Group One included all patients whose follow-up period was at least 24 months; Group Two included those patients with additional follow-up assessments at 36 months. The final samples included 308 and 265 subjects in Groups One and Two, respectively, who met criteria for inclusion. Both Groups included participants who were stable in their diagnosis (MCI-S) and those who converted to a diagnosis of AD over the 2 or 3 years (MCI-C). Table \ref{t1} shows the participant characteristics. Diagnostic criteria for MCI included an MMSE score at baseline between 24 and 30, a CDR score of 0.5, a subjective memory complaint, objective memory loss measured by education-adjusted scores on the Logical Memory II subscale of the Wechsler Memory Scale, generally preserved activities of daily living and no dementia. The diagnostic criteria are an MMSE score between 20 and 26, CDR score of 0.5 and 1.0. The clinical status of each MCI subject was re-assessed at each follow-up visit and updated to reflect one of several outcomes (MCI or AD and other). The MCI-C and MCI-S group designations were based on this follow-up clinical diagnosis and marked as either 1 for MCI-C or 0 for MCI-S in classification study.
\par


\subsection*{Clinical Cognitive Assessment and Genetics data}

\begin{figure}[!h]
\centering
\begin{subfigure}{0.5\linewidth}
  \centering
  \includegraphics[width=\linewidth]{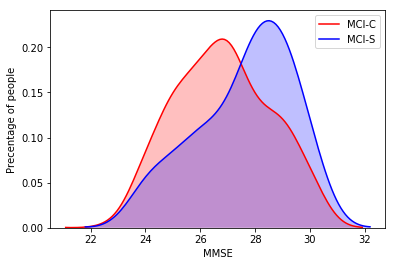}
  \caption{{\bf Distribution of MMSE score in MCI-C and MCI-S groups}
}
  \label{MMSE}
\end{subfigure}%
\begin{subfigure}{0.5\linewidth}
  \centering
  \includegraphics[width=\linewidth]{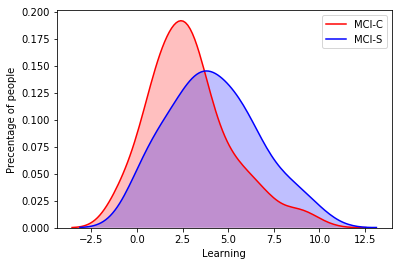}
  \caption{{\bf Distribution of Learning in MCI-C and MCI-S groups}}
  \label{RAVLT}
\end{subfigure}
\caption{{\bf Comparison between MCI-S and MCI-C groups on baseline predictor variables.}
(a) The mean MMSE score in MCI-S is higher than in MCI-C. (b) Mean Learning scores of MCI-C and MCI-S groups are 2.5 and 5.}
\label{MR}
\end{figure}

\begin{figure}[!h]
\centering
\begin{subfigure}{0.475\linewidth}
  \centering
  \includegraphics[width=\linewidth]{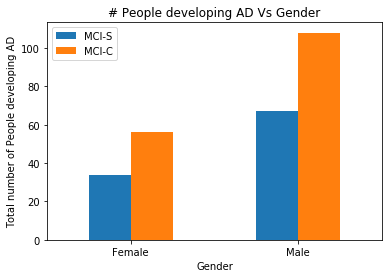}
 \caption{{\bf Sex distribution in the group MCI-S and MCI-C.}
}
  \label{MMSE}
\end{subfigure}%
\hfill
\begin{subfigure}{0.475\linewidth}
  \centering
  \includegraphics[width=\linewidth]{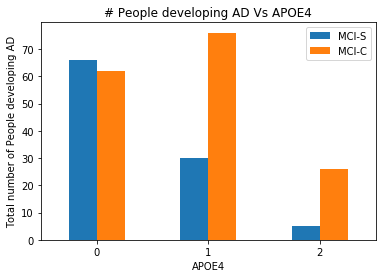}
  \caption{{\bf APOE4 distribution in the group of MCI-S and MCI-C.}
}
  \label{RAVLT}
\end{subfigure}
\vskip\baselineskip
\begin{subfigure}{0.475\linewidth}
  \centering
  \includegraphics[width=\linewidth]{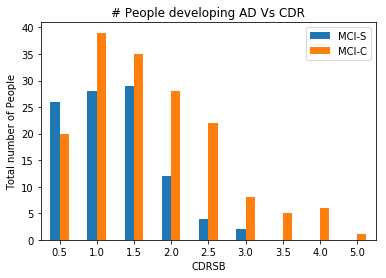}
  \caption{{\bf CDR distribution in the group of MCI-S and MCI-C. }
}
  \label{RAVLT}
\end{subfigure}
\quad
\begin{subfigure}{0.475\linewidth}
  \centering
  \includegraphics[width=\linewidth]{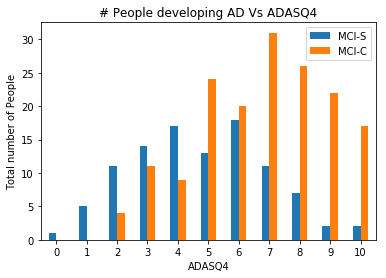}
  \caption{{\bf ASAD distribution in the group of MCI-S and MCI-C.}
}
  \label{RAVLT}
\end{subfigure}
\caption{{\bf Comparison between MCI-S and MCI-C groups on baseline predictor variables.}
 The y-axis of panels (a) through (d) represents the number of participants developing AD. Blue and red bars represent non-converters and converters, respectively. Panel (a) shows a greater number of converters than non-converters for both men and women. Panel (b) shows more than half of MCI-C subjects are APOE4 carriers and approximately 70\% MCI-S subjects are Non-APOE4 carriers. Panel (c) shows MCI-S subjects have the relatively lower CDR score and MCI-C subjects have higher CDR score. The number of people in MCI-C group has a downward trend as CDR score increases. Panel (d) shows MCI-C subjects have the relatively higher ADASQ4 score. The average of ASADQ4 score of MCI-S and MCI-C subjects are approximately around 5 and 8, separately.
}
\label{Comparison}
\end{figure}

We considered a total of 20 clinical features as potential predictors of MCI-to-AD progression in our classification analyses.These included the following assessment scores: the Mini Mental State Examination (MMSE), Clinical Dementia Rating Sum of Boxes (CDR-SB), Alzheimer’s Disease Assessment Scale-cognitive sub-scale (ADAS-cog), Functional Activities Questionnaire (FAQ) measures of activities of daily living, Trail Making Test-B (TRABSCOR), the Rey Auditory Verbal Learning Test (RAVLT), the Digit-Symbol Coding test (DIGT) and the Digit Symbol Substitution Test from the Preclinical Alzheimer Cognitive Composite (mPACCdigit). We also considered genotype for carriers of the epsilon-4 allele of the apolipoprotein E  (APOE) gene ~\cite{bib6} as a genetic predictor in this study. Table \ref{clinical_feature} summarizes all 20 clinical, demographic and genetic features used in this study.
Preliminary comparison of six clinical and genetic predictors by MCI-C and MCI-S subgroups showed five of them (APOE4, ADAS4, CDR, MMSE and RAVLT.learning) significantly differ between the groups, whereas one (SEX) does not. Fig \ref{Comparison} and \ref{MR} illustrate the distribution of these predictors for both groups. On the one hand, MCI-C subjects are more cognitively and functionally impaired  at baseline and exhibit more pronounced verbal memory impairment than MCI-S subjects. This is indicated by higher percentage of APOE4, higher ADAS4 and CDR score, and lower average of MMSE and RAVLT.learning score: the average of MMSE and RAVLT scores MCI-C subjects are approximately 28 and 4, they are 26.5 and 3 for MCI-S groups. On the other hand, the MCI-C and MCI-S groups exhibited similar sex distributions. 

\subsection*{MRI data}

Structural MRI data were collected according to the ADNI acquisition protocol using T1-weighted scans (GradWarp, B1 Correction, N3, Scaled)~\cite{bib27}. This was followed by brain extraction for further processing ~\cite{bib29}. A new multi-atlas registration based label fusion method  was applied for region of interest (ROI) segmentation ~\cite{bib28}. Processing for ROI-based volumetric data used in the present study was performed by researchers at the Perelman School of Medicine, University of Pennsylvanian. These data include baseline MRI scans of ADNI1 participants (230 cognitively normal (CN) individuals, 200 AD patients, and 410 MCI patients). MRI scans were automatically partitioned into 145 anatomic ROIs spanning the entire brain. The segmentation method applies a multi-atlas, consensus-based label fusion scheme on template ROIs deformed to subject space. An additional 114 derived ROIs were calculated by combining single ROIs within a tree hierarchy, to obtain volumetric measurements from larger structures ~\cite{bib27}. In total, 259 ROIs were measured and used as potential predictors of MCI-to-dementia progression in this study. One of the goals of this study is to investigate if manually selecting predictors improves a model's performance. Based on the extant literature ~\cite{bib9}, we manually selected 26 out of 259 features as theoretically significant predictors of MCI to dementia progression (Table \ref{roi}).

\section*{Method and Algorithm}

In the following section, we present two classification techniques: binary LR and SVM, to investigate which exhibits better discrimination accuracy in the context of ADNI data. Several metrics for comparing logistic regression and SVM have been reported in the literature. For example, 1) SVM requires fewer variables than logistic regression to achieve an equivalent level of misclassification rate (MCR) ~\cite{bib10,bib13}; 2) SVM is better than LR with microarray expression data ~\cite{bib13}; 3) SVMs have a nice dual form, giving sparse solutions when using the kernel trick; 4) Both can be viewed as taking a probabilistic model and minimizing some cost associated with the misclassification based on likelihood ratio. Therefore, we show LR and SVM share common roots in statistical pattern recognition and compare the testing accuracy for SVM and logistic regression when using  multi-modal ADNI data.

\subsection*{Logistic Regression}

Logistic regression (LR) is the most commonly used approach in machine learning for binary classification. In the past decade this has been applied to task of AD conversion ~\cite{bib5,bib8,bib35}. We consider a supervised learning task where we are given M training examples $\{D= (x_{i}, y_{i}), i=1,...M\}$. Here each $x_{i} \in \Re^N$ is $N$ dimensional feature vectors, and $y_{i} \in \{0,1\}$ is a class label. The goal of LR is to model the probability $p$ of a random variable $\bm{y}$ being 1 or 0 given the experimental data $\bm{x}$.
Logistic regression models is defined as follows: 
\begin{equation}
    logit \enspace p=log \frac{p}{1-p}
\end{equation}
Logit, the natural logarithm of an odds is the key concept that underlies logistic regression. The formal mathematics equation for logistic regression is: 
\begin{equation}
log \frac{P(y_i=1|x_i;\bm{\beta})}{1-P(y_i=1|x_i;\bm{\beta})}= \sum_{j=1}^N\bm{\beta_{j}}\bm{x_{ij}}
\end{equation}
where $\bm{\beta}=(\beta_1,...\beta_N)^T $ are the parameters or weights of the logistic regression model, $\bm{x_{ij}}=(x_{i1},...x_{iN}$), $i=1,...M$. Also, $P(y_i=1|x_i,\bm{\beta})$ is the probability that $ith$ MCI patient will develop the AD and $P(y_i=0|x_i,\bm{\beta})$ is the probability that $ith$ MCI patient will not develop the AD. Denote $P(y_i=1|x_i;\bm{\beta})=h(x_i)$, then
\begin{equation}
h(x_i)=\frac{1}{1+exp(\sum_{j=1}^N-\bm{\beta_j}x_{ij})}
\end{equation}
Logistic regression is usually trained by minimizing an error function; an appropriate choice of such a function for binary classification problems is the cross-entropy error
\begin{equation}
e_i(\bm{\beta})=-y_ilog(h(x_i))-(1-y_i)log(1-h(x_i)))
\end{equation}
The total cost over the data $ \{D= (x^{i}, y^{i}), i=1,...M\}$ is:
\begin{equation}
J(\bm{\beta})=-\frac{1}{M}[\sum_{i=1}^My_ilog(h(x_i))-(1-y_i)log(1-h(x_i))]
\end{equation}
Consider the problem of finding the maximum likelihood estimate (MLE) of the parameters $\bm{\beta}$ for the unregularized logistic regression model. To find the optimized weights $\bm{\beta}$, the total cost need to be minimized. The optimization function can be written:
\begin{equation}
\bm{\beta}^{optimal}=min_{\bm{\beta}}-\frac{1}{M}[\sum_{i=1}^My_ilog(h(x_i))-(1-y_i)log(1-h(x_i))]\
\label{LR:l1_COST}
\end{equation}
Solving Eq \eqref{LR:l1_COST} yields the optimal weights of $\bm{\beta}$. However, the model-building challenge is to abstract the underlying distribution from the particular instance D of samples because of the small sample compared to the number of features. The problem of memorizing the data set instead of identifying the underlying distribution is known as overfitting~\cite{bib30}. To avoid the overfitting problem, dimension reduction techniques are necessary to be applied. $L_1$ and $L_2$ norm is widely used to avoid over-fitting, especially when there is a only small number of training examples, or a larger number of features to be learned. $L_1$ norm or $lasso$ is also often used for feature selection, and has been shown to have good generalization performance in the presence of many irrelevant features ~\cite{bib16,bib49}. $L_1$ regularization is implemented by adding $L_1$ norm to the cost function, the cost function and the optimization function are following:
\begin{equation}
\label{eqn:LR_L1_P}
J(\bm{\beta})=-\frac{1}{M}[\sum_{i=1}^My_ilog(h(x_i))-(1-y_i)log(1-h(x_i))]+\lambda|\bm{\beta}|
\end{equation}
and
\begin{equation}
\label{eqn:lr_l1}
\bm{\beta}^{optimal}=min_{\bm{\beta}}\{-\frac{1}{M}[\sum_{i=1}^My_ilog(h(x_i))-(1-y_i)log(1-h(x_i))]+\lambda|\bm{\beta}|\}
\end{equation}
where $\lambda$ is positive tuning parameter. This Eq \eqref{eqn:lr_l1} is refereed to as $L_1$ regularized logistic regression.
\subsection*{Support Vector machine}
In this section we discuss the second proposed learning algorithm, Support Vector Machine (SVM), a classification and regression method that can handle high-dimensional feature vectors. Algorithmically, SVMs build optimal boundaries between data sets by solving a constrained quadratic optimization problem ~\cite{bib14,bib15,bib32,bib33,bib34}. Because different kernel functions can be used in the model, we can vary the degrees of non-linearity and flexibility. SVM has received considerable research interest in classification of conversion from MCI to AD over the past years ~\cite{bib1,bib2,bib4,bib6,bib7,bib9,bib31,bib36,bib37,bib46}. \par 
We briefly review a basic support vector machines for classification problems: 
Let $\bm{\beta}^Th(x)+\beta_0=0$ denote an equidistant hyperplane (decision surface) to the closest point of each class on the new space. The goal of SVMs is to find $\bm{\beta}$ and $\beta_0$ such that $|\bm{\beta}^Th(x) + \beta_0| = 1$ for all points closer to the hyperplane. The classifier is constructed as follows. One assumes that: 
\begin{equation}
\bm{\beta}^Th(x_i)+\beta_0=\left\{\begin{array}{ll}
                  \geq1 \; if\; y_i=1\\ 
                  \leq-1\; if \;y_i=0\\
                 \end{array}
                 \right.
\end{equation}
such that the distance from the closest point of each class to the hyperplane is $1/||\bm{\beta}||$ and the distance between the two groups is $2/||\bm{\beta}||$. To maximize the margin, the SVM requires the solution of the following optimization primal problem:

\begin{equation}
\label{eqn:eqlabel}
 min_{\bm{\beta},\bm{\beta_0}} \;\sum_{i=1}^M\{1-y_i[\beta_0+\sum_{j=1}^N\beta_j^Th_j(x_i)]\}
\end{equation}
To make the algorithm work for non-linearly separable data sets as well as be less sensitive to outliers, we reformulate our optimization by adding $L_1$-norm of $\beta$,i.e. the $lasso$ penalty as follows:


\begin{equation}
\label{eqn:svm_l1}
 min_{\bm{\beta},\bm{\beta_0}} \;\sum_{i=1}^M\{1-y_i[\beta_0+\sum_{j=1}^N\beta_j^Th_j(x_i)]\}+\lambda||\beta||_1
\end{equation}
Where $\lambda$ is the tuning parameter that controls the trade-off between loss and penalty. 
\subsection*{Experimental Design}
\begin{figure}[!p]
\centering
\includegraphics[width=1\linewidth]{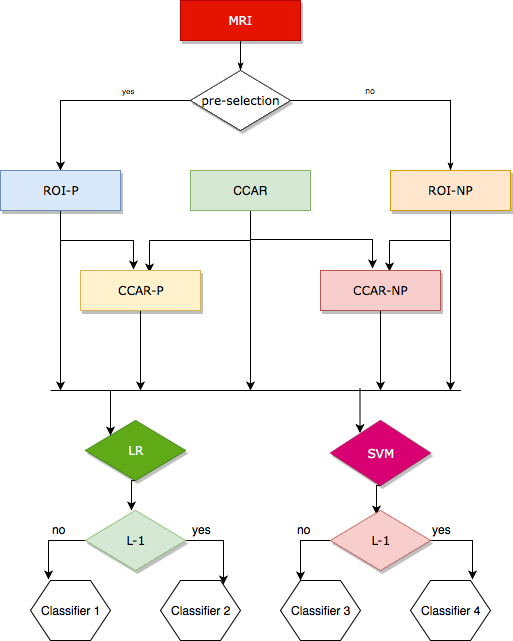}
\caption{{\bf Flowchart of the LR and SVM method}
A) ROI-P:ROI level data with Pre-selection;B) ROI-NP:ROI level data with No Pre-selection;C) CCAR: Clinical, Cognitive assessments score, APOE4 and ROI level data.
}
 \label{classifer}
\end{figure}

We built four different classifiers, each designed to classify individual patients as belonging to either the MCI-C and MCI-S group. Classifier 1 is logistic regression (C-LR); Classifier 2 is logistic regression with $L_1$ norm (C-LR-1); Classifier 3 is support vector machine (C-SVM), and Classifier 4 is SVM with $L_1$ norm (C-SVM-1). To test classifiers' performance on the multi-modal data, we constructed five different data sources,  summarized in Table \ref{Modalities}. First, a separate single data  was constructed for clinical cognitive assessments score and APOE4 (CCA), MRI markers (ROI-NP), MRI markers with pre-selection (ROI-P); Second, a multi-modal data was constructed for the CCAR-NP where joins clinical cognitive assessments score and MRI markers, and CCAR-P which contains the combination of clinical cognitive assessments score and MRI markers with pre-selection. The prediction procedure consisted of three processing stages for Group one (Time=36 months) and Group Two (Time=24 Months): 1) Split data as training and testing set; 2) Train classifiers using training set and assess classifiers using testing set, then train classifiers again using $L_1$ norm on the same training set; 3) Report the testing accuracy, sensitivity and specificity of each classifier on single-modality data. Specifically, in the first stage, $90\%$ used as training set and remaining $10\% $ data are testing set. In the second stage, optimal subsets of features of each data source are determined and chosen when $L_1$ norm was employed. We then list the top 10 features of each data set. In the last stage, we report testing accuracy, sensitivity (percent of MCI-C subjects correctly classified) and specificity (percent of MCI-S subjects correctly classified) as measures of classification accuracy. To protect against over-fitting problem and avoid optimistically-biased estimates of model performance, we report 20 measures of predictive performance for each classifier (1-4) with different partitions of the data, then calculate and report the mean and standard deviation of testing accuracy, sensitivity and specificity. We also investigate the relationship between the number of features and model performance. Finally, we compare the performance of LR with SVM based on their ability to handle the \say{small n, large p} problem as well as the testing accuracy. Fig \ref{classifer} illustrates the diagram of the prediction framework.

\section*{Results and Analysis}
\subsection*{Cross-validation and choice of $\lambda$}
We adopted 10-fold cross-validation to avoid optimistically-biased estimates of model performance. The results of a 10-fold cross-validation run are summarized with the mean and standard deviation of the model skill scores.
Cross-validation are also applied to  tune the hyper-parameters and $\lambda$ is the hyper-parameter as denoted for both LR-$L_1$ and SVM-$L_1$ . To select the optimized $\lambda$, we tried different values of the $\lambda$. Here we report the results with the values of $\lambda=10^{-15},10^{-10},10^{-8},10^{-4},10^{-3},10^{-2},1,5,10,20$ and applied them to the Eq~(\ref{eqn:lr_l1}) and (\ref{eqn:svm_l1}), then we selected the $\lambda$ based on the best cross-validation score and  used selected $\lambda$ to Classifier 2 and 4 to select optimal features. For brevity, the model performance are reported in Tables \ref{result_one} and \ref{result_two} for different Modality, and top 10 selected features are report in Table \ref{select_one_lr}. For example, the best $\lambda$ for ROI-NP-$L_1$ is $0.01$ and the top 3 optimal features selected by LR are  Left Amygdala, Right Accumbens Area and Right MTG middletemporal gyrus.
\subsection*{Comparison with different modalities}
We compared performance of each classifier (1-4) on the five different data sets listed on Table \ref{Modalities}. Performance was gauged using measures of testing accuracy, sensitivity and specificity. As shown in Table \ref{result_one}, the LR with $L_1$ regularization (Classifier 2) can achieve the high accuracy of 85.4\% on multi modality data (CCAR), which is considerably better than performance of LR on the other four modalities. Similarity, the best testing accuracy achieved by SVM is 84.2\% based on the combination of CCAR and SVM-L1. Furthermore, we also found the highest accuracy achieved by both classifiers without applying regularization is based on the single modality data (CCA); this indicated both classifiers perform best on multi-modal when $L_1$ is employed and on single-modality when $L_1$ is not applied. 
\subsection*{Comparison of Pre-selection and $L_1$ norm}
We found that using prior knowledge to inform feature selection improves the model performance and protects against over-fitting. As shown in Table \ref{result_one}, the performance of model on ROI-P $(74.4\%)$ and CCAR-P $(80.4\%)$ outperformed ROI-NP $(63.7\%)$ and CCAR-NP $(64.6\%)$. However, the performance of Classifier 2 on the ROI-NP-$L_1$ and CCAR-NP-$L_1$ have testing accuracy of $85\%$ and $85.4\%$, while the ROI-P-$L_1$ and CCAR-P-$L_1$ have testing accuracy of $79\%$ and $84\%$; this suggests that pre-selection may not have significantly improved model performance compared to $L_1$ norm. In addition to LR, the SVM (Classifiers 3 \& 4) had a similar and comparable result with LR classifiers. First, the observed testing accuracy for CCAR-P and ROI-P are $78.3\%$ and $73.5\%$, which is superior to testing accuracy on CCAR-NP ($72.5\%$) and ROI-NP ($72.3\%$). Therefore, manually selecting features improves model performance only when $L_1$ norm is not applied. Second, it may not necessary to use pre-selection when $L_1$ norm is used, because model on CCAR-NP-$L_1$ with testing accuracy of $84.2\%$ has superior performance over the model on CCAR-P-$L_1$ with testing accuracy of $81.7\%$. 
\subsection*{Comparison of Groups One and Two}
In addition to the results of Group One (i.e., MCI-to-AD conversion over 36 months) , we also reported the performance of Group Two (i.e., MCI-to-AD conversion over 24 months) for comparison. Table \ref{result_two} summarize the predictive performance of LR and SVM for Group Two. Similarity, the performance of Group Two were also investigated for single modality, as well as for multi-modal. The best result is obtained by using LR-$L_1$ model (Classifier 2) on CCAR-NP-$L_1$, and its corresponding testing accuracy is 80.7\%. However, it is warrants mention that all classifiers' performance of Group One outperforms the same classifiers' performance on the same data in group Two. For example, Classifier 2 of Group One on ROI-NP can achieve the testing accuracy of 85\%, which is considerably better than the same classifier of Group Two on ROI-NP (76.3\%); classifier 3 of Group One on CCAR-P had a testing accuracy of 84\% while a testing accuracy on CCAR-P is 79.3\% for Group Two. The experimental results indicated that model performance on data obtained using long follow-up period is better than using short follow-up period. Given the uncertainty in conversion, a longer time window for assessment of change clearly yields more accurate classification.
\subsection*{Comparison of LR and SVM}
\begin{figure}[!p]
\centering
\begin{subfigure}{0.5\linewidth}
  \centering
  \includegraphics[width=\linewidth]{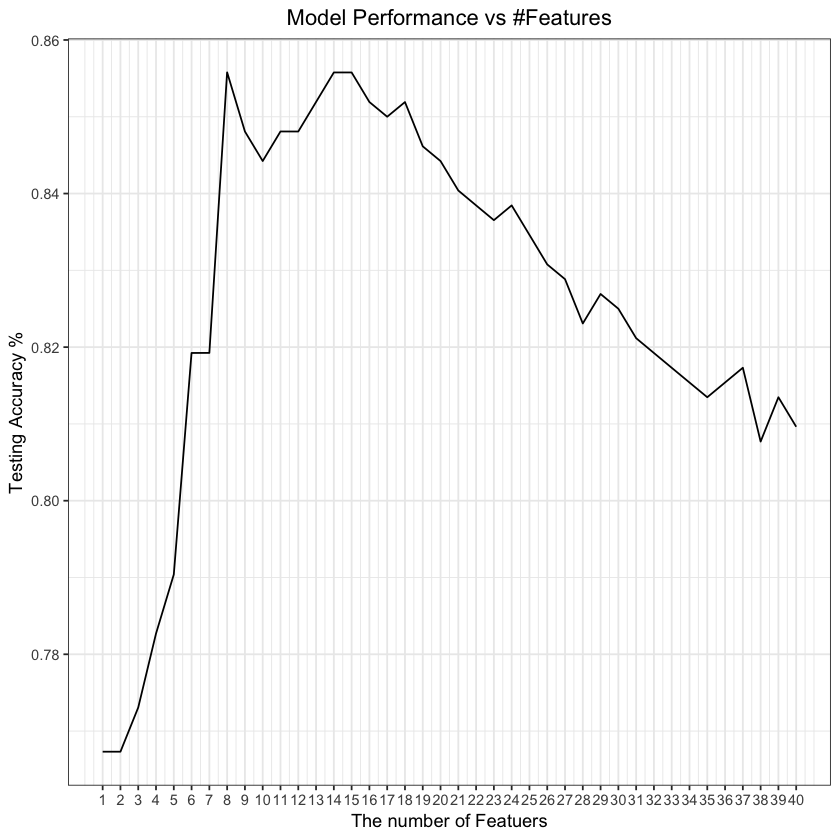}
  \caption{{\bf LR performance on ROI vs the number of features}}
  \label{}
\end{subfigure}%
\begin{subfigure}{0.5\linewidth}
  \centering
  \includegraphics[width=\linewidth]{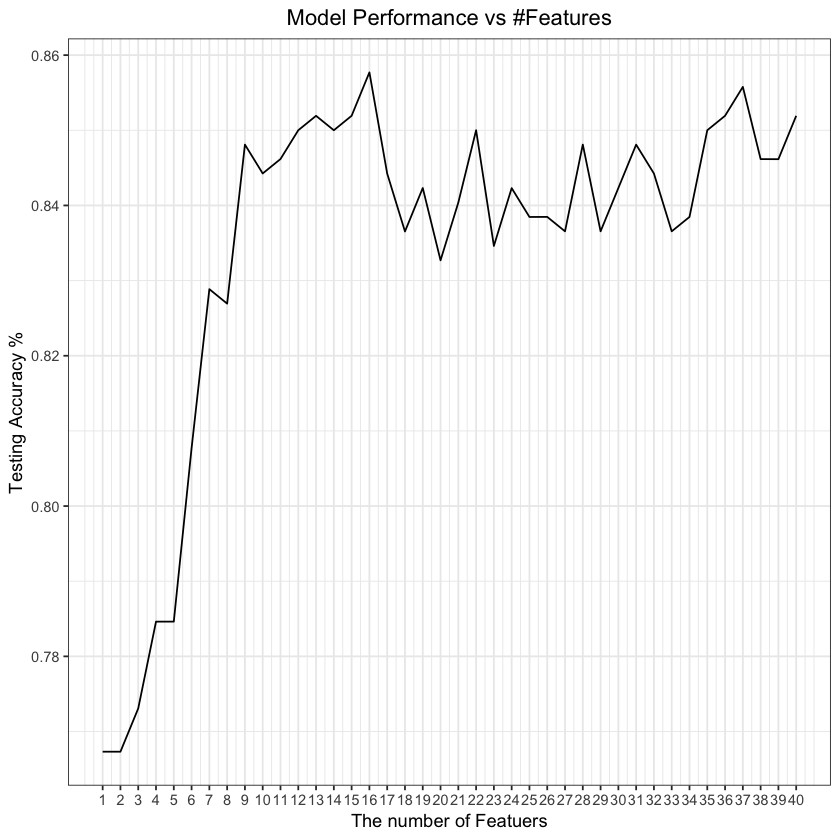}
  \caption{{\bf SVM performance on ROI vs the number of features}}
  \label{}
\end{subfigure}
\caption{
Panel (a) shows dramatic growth in testing accuracy with LR as the number of features increases from 1 to 9; then testing accuracy remained fairly static at approximately 85\% as the number of features increase from 9 to 15. Finally, the testing accuracy drops significantly when the number of features reaches to 16. Panel (b) shows the testing accuracy increased up dramatically as the number of features changes from 1 to 15, but fluctuated after 16. The optimal number of ROI features for both methods are 9 and 15, and their corresponding optimized testing accuracy were approximately 85\%.}
\label{roi_lr_svm}
\end{figure}

\begin{figure}[!p]
\centering
\begin{subfigure}{0.5\linewidth}
  \centering
  \includegraphics[width=\linewidth]{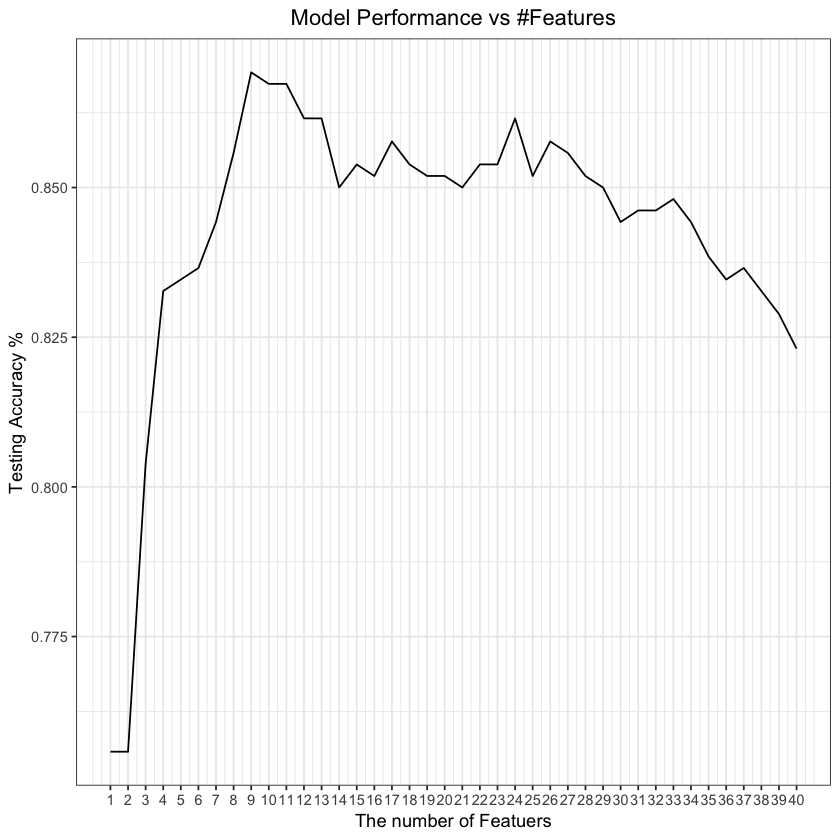}
  \caption{{\bf LR performance on CCAR vs the number of features}}
  \label{}
\end{subfigure}%
\begin{subfigure}{0.5\linewidth}
  \centering
  \includegraphics[width=\linewidth]{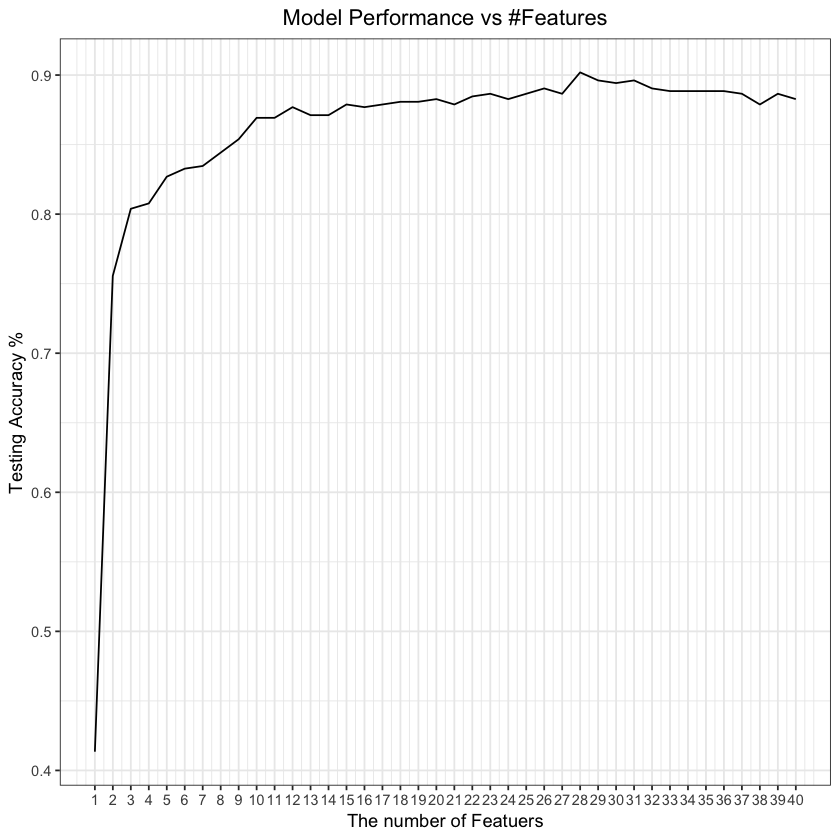}
  \caption{{\bf SVM performance on CCAR vs the number of features}}
  \label{}
\end{subfigure}
\caption{
Figure (a) shows there is a significant increase in the testing accuracy with LR as the number of features increases from 1 to 12, then the testing accuracy fell gradually when the number of features is greater than 12. Figure (b) shows there is  a dramatic growth in the testing accuracy as the number of features increases from 1 to 10, then  testing accuracy remained fairly static at approximately 85\% as the number of features increases. The optimal number of CCAR features for both methods are 9 and 25, and their corresponding optimized testing accuracy are approximately 86\% and 90\%.}
\label{ccar_lr_svm}
\end{figure}

\begin{figure}[!p]
\centering
\begin{subfigure}{0.5\linewidth}
  \centering
  \includegraphics[width=\linewidth]{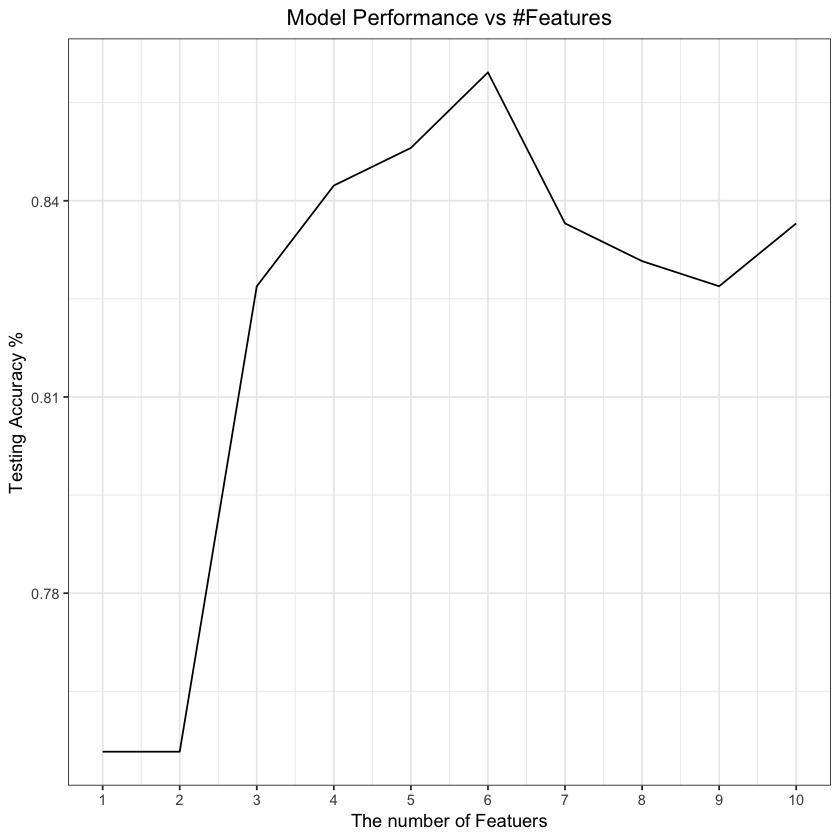}
  \caption{{\bf LR performance on CCA vs the number of features}}
  \label{}
\end{subfigure}%
\begin{subfigure}{0.5\linewidth}
  \centering
  \includegraphics[width=\linewidth]{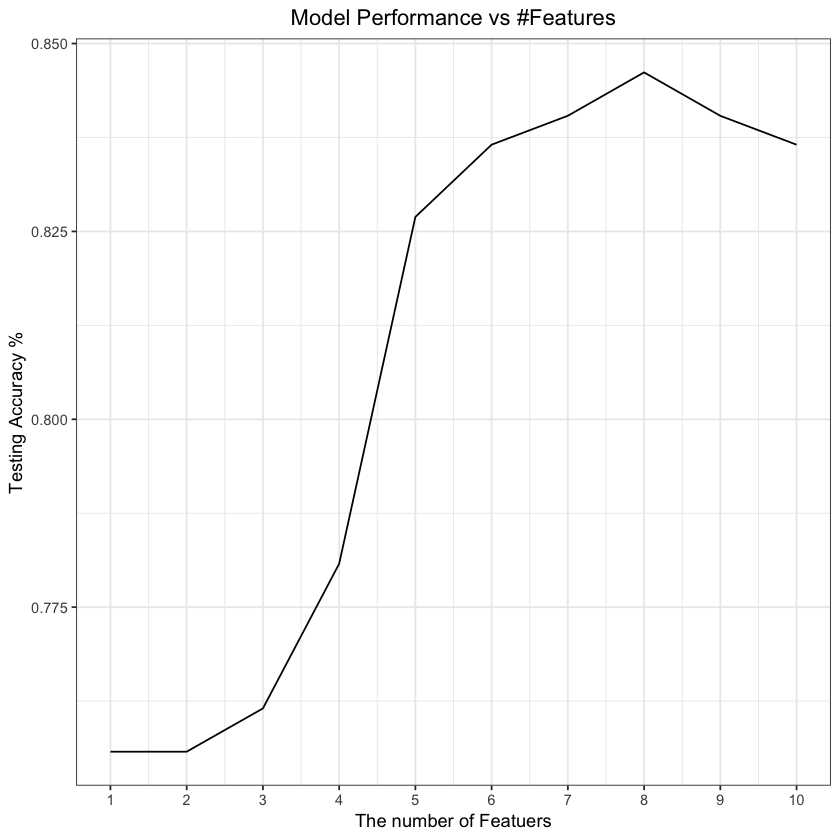}
  \caption{{\bf SVM performance on CCA vs the number of features}}
  \label{}
\end{subfigure}
\caption{
Figure (a) shows there is a significant increase in the testing accuracy with LR as the number of features increases from 1 to 6, then there is a slight decrease in the testing accuracy when the number of features is greater than 6. Figure (b) shows the testing accuracy shot up dramatically as the number of features increases from 1 to 8. The optimal number of CCA features obtained by LR and SVM are 6 and 8, and their corresponding optimized testing accuracy are approximately 85\% and 84.5\%.}
\label{cca_lr_svm}
\end{figure}

In addition to comparing classification between different time windows of assessment, we also compared the differences in performance between LR and SVM. The results, including models’ ability to address the over-fitting problem of LR and SVM methods with different modalities are displayed in Table \ref{result_one}, \ref{result_two} and Fig.\ref{roi_lr_svm}, \ref{ccar_lr_svm} and \ref{cca_lr_svm}. 
First, it is worth noting that both LR and SVM do not work well if no $L_1$ penalization used, since Classifier 2 and 4 outperform Classifier 1 and 3 on the same data set.  Second, as testing accuracy of Classifier 3 on ROI-NP and CCAR-NP are $72.5\%$ and $72.3\%$, both of which outperform classifier 1 on the same data set, it is worth noting that SVM has a better performance on MRI data when feature selection method is not employed. Third, it was interesting to note that we can obtain a good testing accuracy only using simplest model such as LR and the model performance is as good as SVM for a \say{large n small p} data (ROI-P), because testing accuracy of Classifier 1 and 3 on ROI-P are $74.4\%$ and $73.5\%$. Finally,
as shown in Fig. \ref{cca_lr_svm}, \ref{roi_lr_svm}, and \ref{ccar_lr_svm}, the SVM method is more stable and robust than LR to the large number of features when n is small. To summarize, the best performance of group One was achieved by Classifier 2 (LR with $L_1$ norm) when using a combination of three modalities, i,e., CCAR-NP-L1, which has a testing accuracy of $85.4\%$.

\section*{Discussion and Conclusion}
In this study, we applied two machine learning methods to data from clinical assessment and cognitive testing, demographic and genetic markers, and MRI-based volumetric brain measures from ADNI to test classification accuracy in discriminating patients with MCI who progress to AD (MCI-C) from whose who remain stable (MCI-S). We compared LR and SVM classification accuracy and pre-selection dimensional reduction techniques, both as informed by clinical neuroscience expertise and via $L_1$ norm, yielding multiple noteworthy findings. First, the classification results showed that the model using multi-modal data with cognitive, clinical, and volumetric data (CCAR) achieved better classification accuracy than the methods based on single-modality (CCA, ROI). Moreover, the testing accuracy of CCAR based on LR or SVM was either statistically significantly or at least numerically greater than those based on the single-modality model. These results were consistent with the previously published reports ~\cite{bib1,bib2,bib4,bib5,bib6,bib7,bib8,bib9}. We reported the highest testing accuracy on CCAR data is 85.4\% by $L_1$ LR and 84.2\% by $L_1$ SVM. 
Second, SVM has several advantages in discriminating MCI-C and MCI-S (Fig. \ref{roi_lr_svm} and \ref{ccar_lr_svm}). The performance of the SVM approaches was more stable than LR when the number of features was relatively large. In other words, the testing accuracy of SVM on both ROI and CCAR data remained more stable than LR. To be specific, there is a dramatic growth in SVM performance on ROI and CCAR data as the number of features increases from 1 to 12 and from 1 to 10 separately. Testing accuracy remained fairly static at approximately 85\% for both data sets despite the number of features continue to increase. However, LR model performance decreases gradually after the number of ROI features reaching to 12 and CCAR features reaching to 9. 
Third, the classification results proved that manually selecting features on MRI data not only improved the testing accuracy and protected the classifier from overfitting, but also affords easier interpretation of each selected features’ contribution to the model. The reported testing accuracy on ROI-P from LR and SMV are 74.4\% and 73.5\%, compared to testing accuracy on ROI-NP are 63.7\% and 72.5\%. In addition, even though the pre-selection improves the performance of model, Tables \ref{result_one} and \ref{result_two} suggest it is not the best strategy to obtain the maximum testing accuracy, compared to L-1 features selection. As listed in table \ref{result_one}, the testing accuracy of LR on CCAR-P and CCAR-NP-$L_1$ are 80.4\% and 85.4\%. \par
Several prior studies between 2010 and 2016 were conducted to predict the conversion of MCI to AD, using either SVM or LR, and with combining of different modalities that range from MRI, fMRI, PET to CSF ~\cite{bib1,bib2,bib4,bib5,bib6,bib7,bib8,bib9,bib22}. For example, in Young et al (2013) work ~\cite{bib6}, the authors used both SVM and Gaussian process (GP) classification on the MRI, PET, APOE4, and CSF biomarker data from  ADNI. In contrast with other published work, they trained classifier on healthy people and AD patients instead of on group of MCI-C and MCI-S and tested classifier on group of MCI-C and MSI-S. They reported that the accuracy using GP is substantially higher than using any individual modality or using a multi kernel SVM. The reported accuracy is 74\%. In 2016, Korolev et al. implemented multiple kernel learning (pMKL) classification techniques using clinical, MRI and plasma biomarkers data. Their methodology was to first to group the data set into five different data sources, and then apply a filter-wrapper approach of feature selection techniques which combined to use with Joint Mutual Information (JMI) criterion to identify the important features. They report the testing accuracy was 80\%  ~\cite{bib9}.
Most recent computational neuroimaging studies in the past few years have utilized multi-modal features ~\cite{bib4,bib31,bib35,bib36,bib37,bib38,bib39,bib40,bib42,bib46}. In 2017, Ding implemented and applied SVM method on PET and MRI data to classify the transition from MCI to AD, they reported the testing accuracy is 66.35\% ~\cite{bib31}. In addition to use PET and MRI data, other findings show that cerebrospinal fluid (CSF) protein markers can be used to predict progression from MCI to AD, in addition to proteomic, demographic and cognitive data ~\cite{bib43,bib44,bib45}. Llano et al in 2017 applied LR with $L_1$ norm using CSF markers to classify individual patients as belonging to either the MCI-C and MCI-S group, the proposed method obtained a testing accuracy of 70\% ~\cite{bib35}.\par
In conclusion, for the prediction of MCI-to-AD conversion, prior knowledge on feature selection may protect the model against over-fitting, and models with pre-selection on MRI data outperform those without pre-selection. In addition, the present findings demonstrate that SVM classifier performance is more stable than LR for dealing with the problem of \say{small n, large p}. Clinical researchers should note the need to evaluate different classification and pre-selection approaches in application to clinical or research questions, and heed the cautionary note that not all machine learning is equally beneficial to specific clinical outcomes.

\section*{Acknowledgments}
Data used in preparation of this article were obtained from the Alzheimer’s Disease Neuroimaging Initiative (ADNI) database (http://adni.loni.usc.edu). As such, the investigators within the ADNI contributed to the design and implementation of ADNI and/or provided data but did not participate in analysis or writing of this report. A complete listing of ADNI investigators can be found at: \url{http://adni.loni.usc.edu/wp-content/uploads/how_to_apply/ADNI_ Acknowledgement_List.pdf}. Michael W. Weiner, MD is the Principal Investigator of the ADNI; email: Michael.Weiner@ucsf.edu. We are grateful to the patients and their families who participated in the ADNI.The research is partially supported by NSF-DMS 1945824 and 1924724.
\section*{Conflicts of interest}
None of the authors has any conflicts to report.


\newpage 

\begin{table}[h]
\caption{The number of people in Group One and Two}
\centering
\scalebox{1.3}{
\begin{tabular}{|ccccc|}
\hline 
Group &Time & \# MCI-S (y=0) & \# MCI-C (y=1) & \# Total patients  \\ \hline 
One & 36 months & 101 & 164 & 265 \\ \hline
Two & 24 months & 122 & 186 &  308 \\ \hline
\end{tabular}
}
\label{t1}
\begin{tablenotes}
      \small
      \item Table shows the number of MCI-C, MCI-S and total subjects in Group One and Two. The number of MCI-C patients is higher than MCI-S patients in both groups. 
    \end{tablenotes}
\end{table}

\begin{table}[h]
\centering
\caption{Clinical Features and Cognitive Assessment Score of Group One}
\scalebox{1.25}{
\begin{tabular}{|ccccc|}
\hline 
Characteristics& MCI-S & MCI-C & Test statistic & P-value \\ \hline  
Age(years) & $74.34\pm 7.78 $ & $74.84\pm 6.83 $  & -0.528 & $>0.5^a$\\
\hline 
Education(years) & $15.57\pm2.94  $ & $15.73 \pm 2.91 $  & -0.527 &$>0.5^b$ \\
\hline
Sex, $ \%$ female & $33.67\%$ & $34.14\%$ & 0 &$1^b$\\
\hline
APOE4 carriers $ \%$ & $34.65\%$ & $62.19\%$ & 17.900 &$<0.001^a$\\
\hline
CDRSB  & $1.23 \pm 0.61 $ & $1.72 \pm 0.92$ & -5.237 &$<0.001^a$\\
\hline
MMSE score & $27.61 \pm1.74  $& $26.82\pm 1.71  $ &  3.645 &$<0.001^a$\\
\hline
ADAS11 & $8.89 \pm3.79  $& $12.29 \pm 4.16 $ &  -6.823 &$<0.001^a$\\
\hline
ADAS13 & $14.48 \pm 5.50 $& $20.01 \pm 5.79  $ &  -7.795 &$<0.001^a$\\
\hline
ASASQ4 & $4.76 \pm 2.19  $& $6.77 \pm 2.21 $ &  -7.339 &$<0.001^a$\\
\hline
RAVLT.immediate & $36.21\pm 10.10 $& $29.10 \pm 7.98 $ &  6.021 &$<0.001^a$\\
\hline
RAVLT.learning & $4.19\pm2.47  $& $2.91 \pm 2.26  $ &  4.231 &$<0.001^a$\\
\hline
RAVLT.forgetting  & $4.31\pm2.59  $& $4.47 \pm 2.15 $ &  -1.501 &$0.135^a$\\
\hline
RAVLT.perc.forgeting  & $51.55 \pm31.04  $& $72.85 \pm 30.45 $ &   -5.464 &$<0.001^a$ \\
\hline
LEDLTOTAL  & $4.96 \pm2.36 $& $3.41 \pm 2.66 $ &  4.931 &$<0.001^a$\\
\hline
DIGTSCOR & $40.75 \pm 11.09 $& $36.72 \pm 10.96 $ &  2.883 &$<0.005^a$\\
\hline
TRABSCOR & $109.43\pm62.94 $& $132.09\pm 71.36 $ &  -2.704 &$0.007^a$\\
\hline
FAQ & $1.50 \pm2.99 $& $4.96 \pm 4.79$ & -7.243 &$<0.001^a$ \\
\hline
mPACCdigit & $-5.376\pm2.96  $& $-8.06\pm 2.96 $ &  7.174 &$<0.001^a$\\
\hline
mPACCtrailsB & $-5.47 \pm3.06$& $-8.22 \pm2.98  $ & 7.174 &$<0.001^a$\\
\hline
\end{tabular}
}
\label{clinical_feature}
\begin{tablenotes}
\small{
\item Table only for Group one where has 265 patients and 36 months follow-up time. Values are shown as mean $\pm$ standard deviation or percentage. Test statistics and P-values for differences between MCI-S and MCI-C are based on (a) t-test or (b) chi- square test. MCI-S = non-progressive MCI; MCI-P = progressive MCI; APOE = apolipoprotein E; MMSE = Mini-Mental State Examination. RAVLT = The Rey Auditory Verbal Learning Test (immediate: sum of 5 trails; learning: trial 5-trial 1; Forgetting: trial 5-delayed; perc.forgetting: Precent forgetting) ; DIGT = The Digit- Symbol Coding test; TRAB = Trail Making tests; CDRSB = Clinical Dementia Rating Scaled Response; FAQ = Activities of Daily living Score; ADAS = Alzheimer's Disease Assessment Scale–Cognitive sub- scale; mPACCdigit = the Digit Symbol Substitution Test from the Preclinical Alzheimer Cognitive Composite;}
\end{tablenotes}
\end{table}

\begin{table}[h]
\centering
\caption{Pre-selected MRI Features of Group One}
\scalebox{1.3}{
\renewcommand{\arraystretch}{0.6}%
\begin{tabular}{|ccccc|}
\hline
Characteristics& MCI-S & MCI-C &Test statistic &P-value\\ \hline 

HippoR & $3684 \pm 438 $ & $ 3366 \pm 437$ &  5.735&  $<0.001$\\
\hline
HippoL & $3414 \pm 418 $ & $3105 \pm 388$  &  5.994&  $<0.001$\\
\hline
flWMR & $96720 \pm 6218$ & $96976 \pm 5585$   &  -0.338  & 0.73\\
\hline
flWML & $93671 \pm 5836$ & $94238 \pm 5160 $&  -0.802  & 0.42\\
\hline
plWMR & $47197 \pm 3415 $& $47141 \pm 3098 $ & 0.135& 0.89\\
\hline
plWML & $50149 \pm 3714 $& $50038 \pm 3467$ &  0.242& 0.81\\
\hline
tlWMR  & $56076 \pm 3252 $& $55934 \pm 2931 $ & 0.359& 0.72\\
\hline
tlWML  & $55412 \pm 3396 $& $55468 \pm 3023 $ & -0.136& 0.89\\
\hline
ACgCR  & $3167 \pm 756 $& $3128 \pm 641 $ & 0.438& 0.66\\
\hline
ACgCL  & $4104 \pm 787 $& $4075 \pm 689 $ & 0.312& 0.76 \\
\hline
EntR  & $2189 \pm 365 $& $1983 \pm 373 $ &  4.412& $<0.001$\\
\hline
EntL  & $2050 \pm 399 $& $1844 \pm 356 $ &  4.240& $<0.001$\\
\hline
MCgCR  & $4176 \pm 547 $& $4200 \pm 541 $ & -0.341& 0.73\\
\hline
MCgCL  & $3988 \pm 493 $& $4002 \pm 559 $ & -0.213& 0.83\\
\hline
MFCR  & $1581 \pm 342 $& $1505 \pm 524 $ & 1.805& 0.07\\
\hline
MFCL  & $1566 \pm 285 $& $1548 \pm 291 $ & 0.487& 0.62\\
\hline
OpIFGR  & $2575 \pm 608 $& $2425\pm 546 $ & 2.021& 0.04\\
\hline
OpIFGL  & $2465 \pm 550 $& $2361 \pm 579 $ &  1.466&0.14\\
\hline
OrIFGR  & $1252 \pm 315 $& $1196 \pm 362 $ &  1.322& 0.18\\
\hline
OrIFGL  & $1514 \pm 335 $& $1398 \pm 356 $ &  2.658& $<0.001$\\
\hline
PCgCR  & $3679 \pm 466 $& $3528 \pm 415 $ &  2.657&$<0.001$\\
\hline
PCgCL & $3991 \pm 442 $& $3789 \pm424 $ &  3.676& $<0.001$\\
\hline
PCuR  & $10129 \pm 1193 $& $9862 \pm 1313 $ & 1.701& 0.09\\
\hline
PCuL  & $10005 \pm1263 $& $9759 \pm 1299 $ & 1.522& 0.13\\
\hline
SPLR  & $8867 \pm1140 $& $8693 \pm 1219 $ & 1.180& 0.02\\
\hline
SPLL  & $8880 \pm1192 $& $8662 \pm 1313 $ & 1.390& 0.17\\

\hline
\end{tabular}
}
\label{roi}
\begin{tablenotes}
      \item Values are shown as mean $\pm$ standard deviation or percentage. Test statistics and P-values for differences between MCI-C and MCI-S are based on  t-test. MCI-S = non-progressive MCI; MCI-C = progressive MCI. HippoR = Right Hippocampus; HippoL = Left Hippocampus; flWMR = frontal lobe WM right; flWML = frontal lobe WM left; plWMR = parietal lobe WM right; plWML = parietal lobe WM left; tlWMR = temporal lobe WM right; tlWML = temporal lobe WM left; ACgCR=Right ACgG  anterior cingulate gyrus; ACgCL=Left ACgG  anterior cingulate gyrus; EntR = Right Ent entorhinal area; EntL = Left Ent entorhinal area; MCgCR = Right MCgG  middle cingulate gyrus ;MCgCL = Left MCgG  middle cingulate gyrus; MFCR = Right MFC   medial frontal cortex; MFCL = Left MFC   medial frontal cortex; OpIFGR = Right OpIFG opercular part of the inferior frontal gyrus; OpIFGL = Left OpIFG opercular part of the inferior frontal gyrus; OrIFGR = Right OrIFG orbital part of the inferior frontal gyrus; OrIFGL = Left OrIFG orbital part of the inferior frontal gyrus; PCgCR = Right PCgG  posterior cingulate gyrus ; PCgCL = Left PCgG  posterior cingulate gyrus; PCuR = Right PCu   precuneus; PCuL = Left PCu   precuneus; SPLR  = Right SPL superior parietal lobule; SPLL  = Left SPL superior parietal lobule.
    \end{tablenotes}
\end{table}

\clearpage

\begin{table}[!h]
\begin{adjustwidth}{-0.25in}{0in} 
\centering
\caption{Modalities}
\renewcommand{\arraystretch}{1.5}%
\begin{tabular}{|p{12cm}p{1.2cm}|}
\hline
\multicolumn{1}{|c}{\bf Multi-modal data} & \multicolumn{1}{c|}{\bf \# features}\\ 
\hline
Clinical Cognitive Assessments score and APOE4 data(CCA)& 20\\

ROI with no pre-selection data(ROI-NP) & 260\\

ROI with pre-selection data (ROI-P) & 25 \\

CCA, APOE4 and ROI with no pre-selection data (CCAR-NP)& 279\\

CCA, APOE4 and ROI with pre-selection data (CCAR-P) & 44\\
\hline
\end{tabular}
\label{Modalities}
\end{adjustwidth}
\end{table}

\clearpage

\begin{table}[!h]
\begin{adjustwidth}{0in}{0in} 
\centering
\caption{
{\bf Top 10 features of Group One obtained by $L_1$ regularization }}
\scalebox{1}{
\renewcommand{\arraystretch}{1}%
\begin{tabular}{|ccccccc|}
\hline
\multicolumn{1}{|c}{\bf Source} &\multicolumn{3}{c}{\bf LR-L1(Classifier 2)} & \multicolumn{3}{c|}{\bf SVM-L1 (Classifier 4)}
\\ 
Data& CCA  & ROI-NP & CCAR-NP &CCA &ROI-NP & CCAR-NP\\ \hline
1 & FAQ & AmyL & FAQ & FAQ & AmyL & FAQ \\ \hline
2 & mPACCtrailsB & AccmR & AmyL & Yrs. Educ. & AccmR & AmyL  \\ \hline
3 & APOE4 & MTGR & ADASQ4 & APOE4 & AOrGL & AccmR \\ \hline
4 & ADASQ4 & HippoL & HippoL & mPACCdigit & PCgGL &  AOrGL \\ \hline
5 & Learning & AOrGL & MTGR	& ADASQ4 & HippoL & PTR   \\ \hline
6 & Yrs. Educ. & PrGR & APOE4 & Learning & PrGR & AnGR  \\ \hline
7 & Forgetting  & PCgGL & AOrGL & ADAS11 & POrGR & APOE4\\ \hline
8 & mPACCdigit & InfR & Learning & mPACCtrailsB & PTR & PCgGL \\ \hline
9 & ADAS13 & POR & mPACCtrailsB & DELTOTAL & LOrGL & Learning  \\ \hline
10 & ADAS11 & MOGL & mPACCdigit & Forgetting  & MOrGL &POrGR\\ \hline
\end{tabular}
}
\label{select_one_lr}
\begin{adjustwidth}{-0.1in}{-0in} 
\begin{tablenotes}
      \small AccmR = Right Accumbens Area; AmyL = Left Amygdala; HippoL = Left Hippocampus ; InfR  = Right Inf Lat Vent; AOrGL= Left anterior orbital gyrus; AnGR=Left angular gyrus; LOrGL = Left   lateral orbital gyrus; MOGL = Left middle occipital gyrus; MOrGL = Left medial orbital gyrus; MTGR = Right middle temporal gyrus; PCgGL = Left posterior cingulate gyrus;  POR = Right parietal operculum; POrGR = Right posterior orbital gyrus; PrGR = Right precentral gyrus; PTR = Right  planum temporale; 
    \end{tablenotes}
    \end{adjustwidth}
\end{adjustwidth}
\end{table}

\begin{table}[!h]
\begin{adjustwidth}{0in}{0in} 
\centering
\caption{
{\bf LR and SVM performance of Group One (Time = 3 years) for single-data and multi-modal data }}
\scalebox{0.8}{
\begin{tabular}{|cccccccc|}
\hline
\multicolumn{1}{|c}{\bf Source} &\multicolumn{3}{c}{\bf LR (Classifier 1 and 2)} & \multicolumn{3}{c}{\bf SVM (Classifier 3 and 4)}& \multicolumn{1}{c|}{\bf Features}
\\ 
Modality  & Testing Acc $\%$ & Sp $\%$ & Sn $\%$ & Testing Acc $\%$ & Sp $\%$ & Sn $\%$ & \# Features  \\ \hline
CAA & $83.8 \pm 6.56$ & $67.3 \pm 24.8 $ & $91.6 \pm 9.61$ & $81.9 \pm 5.45$ & $66 \pm 17.8 $ & $89.9 \pm 8.43$ & \diagbox[width=6em]{$20^{(1)}$}{$20^{(2)}$} \\ \hline
ROI-NP & $63.7 \pm 9.04$ & $6.41 \pm 19.7 $ & $94.6 \pm 16.7$ & $72.3 \pm 7.45$ & $43.1 \pm 26.4$ & $86.3 \pm 16.5$ & \diagbox[width=6em]{$260^{(1)}$}{$260^{(2)}$}\\ \hline
ROI-P & $74.4 \pm 7.2$ & $41.0 \pm 21.0 $ & $91.1 \pm 12.2 $ & $73.5 \pm 6.82$ & $42.1 \pm 22.3 $ & $88.7 \pm 12.0$ & \diagbox[width=6em]{$25^{(1)}$}{$25^{(2)}$}\\ \hline
CCAR-NP & $64.6 \pm 8.24$ & $11.8 \pm 24.3 $ & $92.8 \pm 14.8$ & $72.5 \pm 7.72$ & $44.7 \pm 28$ & $83.4 \pm 20.8$ & \diagbox[width=6em]{$279^{(1)}$}{$279^{(2)}$}\\ \hline
CCAR-P & $80.4 \pm 7.16$ & $65.6 \pm 25.4 $ & $87.6 \pm 13.0$ & $78.3 \pm 7.2$ & $52.6 \pm 24.3 $ & $92.3 \pm 11.1 $ & \diagbox[width=6em]{$44^{(1)}$}{$44^{(2)}$}\\ \hline
CCA-$L_1$ & $84\pm 7.31$ & $76.3 \pm 18.4$ & $87.8 \pm 12.0 $ & $83.3 \pm 5.89$ & $67.3 \pm 19.6$ & $90.7 \pm 8.72 $& \diagbox[width=6em]{$10^{(1)}$}{$14^{(2)}$}\\ \hline
ROI-NP-$L_1$ & $85\pm 6.59$& $67.9 \pm 20.4$ & $92.8 \pm 5.98 $ & $80.4 \pm 5.84$ & $59.3 \pm 23.8$& $89.2 \pm15.6$ & \diagbox[width=6em]{$17^{(1)}$}{$66^{(2)}$} \\ \hline
ROI-P-$L_1$ & $79 \pm 7.54$ & $50.0 \pm 26.6$ & $92.7 \pm 10.8 $ & $75.8 \pm 6.62$ & $51.3\pm 21.1 $& $88.5 \pm 10.5$ & \diagbox[width=6em]{$5^{(1)}$}{$17^{(2)}$}\\ \hline
CCAR-NP-$L_1$ & $85.4 \pm 4.93$ & $70.7 \pm 20.8 $ & $92.2 \pm 7.03$ & $84.2 \pm 4.98 $& $73.5 \pm 20.5$ & $88.5 \pm  11.1$& \diagbox[width=6em]{$18^{(1)}$}{$64^{(2)}$} \\ \hline
CCAR-P-$L_1$ & $84 \pm 6.39$ & $65.4 \pm 20.6 $& $93.7 \pm 7.12$ & $81.7 \pm 6.10$ & $64.7 \pm 20.3 $ & $90.5\pm 10.6 $& \diagbox[width=6em]{$14^{(1)}$}{$25^{(2)}$}\\ \hline
\end{tabular}
}
\label{result_one}
\begin{adjustwidth}{0in}{0in} 
\begin{tablenotes}
       For each modality, the predictive performance of LR and SVM are shown (mean $\pm$ standard deviation), including testing accuracy, sensitivity (Sn), specificity (Sp),  \# features is the number of features; this parameter was determined via (1): Classifier 2; (2): Classifier 4. 
    \end{tablenotes}
    \end{adjustwidth}
\end{adjustwidth}
\end{table}

\begin{table}[!h]
\begin{adjustwidth}{0in}{0in} 
\centering
\caption{
{\bf LR and SVM performance of Group One (Time = 2 years) for single-data and multi-modal data }}
\scalebox{0.8}{
\begin{tabular}{|cccccccc|}
\hline
\multicolumn{1}{|c}{\bf Source} &\multicolumn{3}{c}{\bf LR (Classifier 1 and 2)} & \multicolumn{3}{c}{\bf SVM (Classifier 3 and 4)}& \multicolumn{1}{c|}{\bf Features}
\\ 
Modality  & Testing Acc $\%$ & Sp $\%$ & Sn $\%$ & Testing Acc $\%$ & Sp $\%$ & Sn $\%$ & \# Features\\ \hline
CAA & $77.8 \pm7.82$ & $62.4 \pm 28.1$ & $86.4 \pm 14.5$ & $78.3 \pm 7.53$ & $65.3 \pm 26.4$ & $84.8 \pm 14.5$ & \diagbox[width=6em]{$20^{(1)}$}{$20^{(2)}$}\\ \hline
ROI-NP & $61.2 \pm 5.95$& $19.1 \pm 30.6$ & $88.8 \pm 19.3$ & $69.7 \pm 5.17$ & $46.5 \pm 26.3$& $83\pm 15.3 $& \diagbox[width=6em]{$260^{(1)}$}{$260^{(2)}$}\\ \hline
ROI-P & $70.2 \pm 6.62$ & $39.4 \pm 25$& $89.9 \pm 11.2$ & $70 \pm 6.75$& $37.3 \pm 26.7 $& $90.8 \pm 13.3 $ & \diagbox[width=6em]{$25^{(1)}$}{$25^{(2)}$}\\ \hline
CCAR-NP & $62.3 \pm 6.13$& $26.7 \pm 33.9$& $83.7 \pm 23.1$& $70.0\pm 5.72 $& $53.4 \pm 27.4$& $78.7 \pm 16.2$ & \diagbox[width=6em]{$279^{(1)}$}{$279^{(2)}$}\\ \hline
CCAR-P & $75.7 \pm 7.42$& $55.8 \pm 27.6 $ & $87.5 \pm 14.4$& $75 \pm 6.07$ & $51.3 \pm 24.6 $& $90.3 \pm 12.3 $& \diagbox[width=6em]{$44^{(1)}$}{$44^{(2)}$} \\ \hline
CCA-$L_1$ & $80 \pm 7.57$& $69.1 \pm 25.3$& $85 \pm 13.4$ & $79.8 \pm 7.37$ & $70.5 \pm 25.6$ & $84 \pm 13.7 $& \diagbox[width=6em]{$12^{(1)}$}{$14^{(2)}$} \\ \hline
ROI-NP-$L_1$ & $76.3 \pm 6.83 $& $55.7 \pm 18.8$ & $90.7 \pm 9.56$ & $76\pm 5.78 $ & $56.6 \pm 20.5$ & $90.5\pm 10.3 $& \diagbox[width=6em]{$19^{(1)}$}{$120^{(2)}$}\\ \hline
ROI-P-$L_1$ & $73.7 \pm 5.5$& $45.8 \pm 23.3$ & $91.8 \pm 8.60$& $71.8 \pm 6.80$& $42.2 \pm 25.3 $& $90.9 \pm 13.5$ & \diagbox[width=6em]{$6^{(1)}$}{$14^{(2)}$}\\ \hline
CCAR-NP-$L_1$ & $80.7 \pm 7.46 $& $67.3 \pm 21.2$ & $89.3 \pm 10.4 $& $80.0 \pm 6.21$& $73 \pm 20.2$& $83.8\pm 11.0 $& \diagbox[width=6em]{$26^{(1)}$}{$79^{(2)}$}\\ \hline
CCAR-P-$L_1$ & $79.3\pm 6.89 $& $65.5 \pm 22.0 $ & $89.1 \pm 7.78 $ & $77.2 \pm 5.44 $ & $58.1 \pm 19.7 $& $89.2 \pm 12.6 $& \diagbox[width=6em]{$14^{(1)}$}{$27^{(2)}$}\\ \hline
\end{tabular}
}
\label{result_two}
\begin{adjustwidth}{0in}{0in} 
\begin{tablenotes}
       For each modality, the predictive performance of LR and SVM are shown (mean $\pm$ standard deviation), including testing accuracy, sensitivity (Sn), specificity (Sp),  \# features is the number of features;  \# features is the number of features; this parameter was determined via (1): Classifier 2; (2): Classifier 4.
    \end{tablenotes}
    \end{adjustwidth}
\end{adjustwidth}
\end{table}

\clearpage

\clearpage

\end{document}